\documentclass[journal]{IEEEtran}

\usepackage{orcidlink}
\usepackage{booktabs}
\usepackage{cite}
\usepackage{amsmath,amssymb,amsfonts}
\usepackage[nolist,nohyperlinks]{acronym}
\usepackage[inline]{enumitem}
\usepackage[detect-weight=true, detect-family=true]{siunitx}
\usepackage{algpseudocode}
\usepackage{algorithm}
\usepackage{svg}
\usepackage{multirow}
\usepackage{balance}
\usepackage{graphicx}
\usepackage[normalem]{ulem}
\usepackage[nameinlink, capitalise]{cleveref}
\usepackage[usestackEOL]{stackengine}
\DeclareSIUnit\ohm{\ensuremath\Omega}
\DeclareSIUnit\db{dB}
\DeclareSIUnit{\belmilliwatt}{Bm}
\DeclareSIUnit{\bel}{B}
\DeclareSIUnit{\bitpersecond}{bps}
\DeclareSIUnit{\samplepersecond}{Sps}
\DeclareSIUnit{\nothing}{\relax}

\makeatletter
\newcommand*{\org@overidelabel}{}
\let\org@overridelabel\@verridelabel
\@ifpackagelater{acronym}{2015/03/21}{
  \renewcommand*{\@verridelabel}[1]{%
    \@bsphack
    \protected@write\@auxout{}{\string\AC@undonewlabel{#1@cref}}%
    \org@overridelabel{#1}%
    \@esphack
  }%
}{
  \renewcommand*{\@verridelabel}[1]{%
    \@bsphack
    \protected@write\@auxout{}{\string\undonewlabel{#1@cref}}%
    \org@overridelabel{#1}%
    \@esphack
  }%
}

\newcommand{\linebreakand}{%
  \end{@IEEEauthorhalign}
  \hfill\mbox{}\par
  \mbox{}\hfill\begin{@IEEEauthorhalign}
}
\makeatother

\definecolor{confcolor}{HTML}{4c72b0}
\usetikzlibrary{calc}

\newcommand{\confmat}[7]{
\begin{tikzpicture}[
    scale = #7,
    ]

\tikzset{vertical label/.style={rotate=90,anchor=east}}   
\tikzset{diagonal label/.style={rotate=45,anchor=north east}}

\def\totSamples{0}
\foreach \y in {1,...,#6}
{
    \foreach \x in {1,...,#6}
    {
      \pgfmathparse{#1[\y-1][\x-1]}   
      \xdef\totSamples{\totSamples+\pgfmathresult} 
    }
}
\pgfmathparse{\totSamples} \xdef\totSamples{\pgfmathresult}  

\foreach \y in {1,...,#6} 
{
    \node [vertical label] at (0.2,-\y+0.4) {\pgfmathparse{#3[\y-1]}\pgfmathresult};

    \foreach \x in {1,...,#6}  
    {
    \begin{scope}[shift={(\x,-\y)}]
        \def\mVal{#1[\y-1][\x-1]} 
        \pgfmathtruncatemacro{\r}{\mVal}   %
        \pgfmathtruncatemacro{\p}{round(\r/\totSamples*100)}
        \coordinate (C) at (0,0);
        \ifthenelse{\p<50}{\def\txtcol{black}}{\def\txtcol{white}} 
        \node[
            draw,                 
            text=\txtcol,         
            align=center,         
            fill=confcolor!\p,        
            minimum size=#7*10mm,    
            inner sep=0,          
            ] (C) {\r\\\p\%};     
        \ifthenelse{\y=#6}{
        \node [] at ($(C)-(0,0.75)$) 
        {\pgfmathparse{#5 [\x-1]}\pgfmathresult};}{}
    \end{scope}
    }
}
\coordinate (yaxis) at (-0.1,0-#6/2);  
\coordinate (xaxis) at (0.5+#6/2, -#6-1.05); 
\node [vertical label] at (yaxis) {#2};
\node []               at (xaxis) {#4};
\end{tikzpicture}
}

\begin{acronym}
  \acro{IoT}{Internet of Things}
  \acro{BLE}{Bluetooth Low Energy}
    \acro{dBN}{dynamic Bayesian Network}
    \acro{KF}{Kalman Filter}
    \acro{PF}{Particle Filter}
    \acro{NI}{non parametric information}
    \acro{BLE}{Bluetooth Low Energy}
    \acro{EKF}{Extended Kalman Filter}
    \acro{RSSI}{received signal strength indicator}
    \acro{LOS}{line-of-sight}
    \acro{UWB}{ultra-wideband}
    \acro{NLOS}{non-line-of-sight}
    \acro{RSME}{root mean squared error}
    \acro{MAE}{mean average error}
    \acro{ToF}{time of flight}
    \acro{MCPD}{multi-carrier phase-difference}
    \acro{RF}{random forest}
    \acro{SoC}{system-on-chip}
    \acro{DL}{deep learning}
    \acro{FFT}{fast Fourier transform}
    \acro{CA}{cumulative average}
    \acro{IQR}{interquartile range}
    \acro{LTE-M}{Long-Term Evolution Machine Type Communication}
    \acro{GNSS}{global navigation satellite system}
    \acro{MQTT}[\textsc{MQTT}]{\textsc{MQTT}}
    \acro{RAM}{random-access memory}
    \acro{AWS}{Amazon Web Services}
    \acro{IMU}{inertial measurement unit}
    \acro{CPU}{central processing unit}
    \acro{RTLS}{real-time location systems}
    \acro{RFID}{radio frequency identification}
    \acro{PDA}{personal digital assitant}
    \acro{PC}{personal computer}
    \acro{ML}{machine learning}
    \acro{MCU}{microcontroller}
\end{acronym}

\usepackage{eso-pic}
\usepackage{url}
\AddToShipoutPictureBG*{
  \AtPageUpperLeft{%
    \put(0,-30){\raisebox{15pt}{\makebox[\paperwidth]{\begin{minipage}{21cm}\centering
      \textcolor{gray}{This article has been accepted for publication in the Internet of Things Journal.\\DOI: 10.1109/JIOT.2024.3479458} 
    \end{minipage}}}}%
  }
  \AtPageLowerLeft{%
    \raisebox{25pt}{\makebox[\paperwidth]{\begin{minipage}{21cm}\centering
      \textcolor{gray}{ \copyright 2024 IEEE.  Personal use of this material is permitted.  Permission from IEEE must be obtained for all other uses, in any current or future media, including reprinting/republishing this material for advertising or promotional purposes, creating new collective works, for resale or redistribution to servers or lists, or reuse of any copyrighted component of this work in other works.
      }
    \end{minipage}}}%
  }
}

\begin{document}


\title{A Proximity-based Approach for Dynamically Matching Industrial Assets and Their Operators Using Low-Power \acs{IoT} Devices}





\author{
    \IEEEauthorblockN{
        Silvano Cortesi\orcidlink{0000-0002-2642-0797}, \textit{Graduate Student Member, IEEE},
        Michele Crabolu\orcidlink{0000-0003-4308-2432},
        Prodromos-Vasileios Mekikis\orcidlink{0000-0003-0616-0657},\\ 
        Giovanni Bellusci\orcidlink{0000-0002-5514-2503},
        Christian Vogt\orcidlink{0000-0003-4551-4876}, \textit{Member, IEEE}} and 
        Michele Magno\orcidlink{0000-0003-0368-8923}, \textit{Senior Member, IEEE}

\thanks{\textit{(Corresponding author: Silvano Cortesi, silvano.cortesi@pbl.ee.ethz.ch.)}}
\thanks{Silvano Cortesi, Christian Vogt and Michele Magno are with the Center for Project-Based Learning, ETH Zürich, 8092 Zürich, Switzerland (e-mail: firstname.lastname@pbl.ee.ethz.ch)}
\thanks{Michele Crabolu, Prodromos-Vasileios Mekikis and Giovanni Bellusci are with Hilti AG, 9494 Schaan, Liechtenstein (e-mail: firstname.lastname@hilti.com)}
}

\markboth{IEEE INTERNET OF THINGS JOURNAL, VOL. xx, NO. xx, xx XXXXXXX xxxx}%
{S. Cortesi \MakeLowercase{\textit{et al.}}}


\maketitle

\begin{abstract}
    Asset tracking solutions have proven their significance in industrial contexts, as evidenced by their successful commercialization (e.g., Hilti On!Track). However, a seamless solution for matching assets with their users, such as operators of construction power tools, is still missing. By enabling asset-user matching, organizations gain valuable insights that can be used to optimize user health and safety, asset utilization, and maintenance. This paper introduces a novel approach to address this gap by leveraging existing \ac{BLE}-enabled low-power \ac{IoT} devices. The proposed framework comprises the following components:
\begin{enumerate*}[label=\roman*)]
\item a wearable device,
\item an \ac{IoT} device attached to or embedded in the assets,
\item an algorithm to estimate the distance between assets and operators by exploiting simple \ac{RSSI} measurements via an \ac{EKF}, and
\item a cloud-based algorithm that collects all estimated distances to derive the correct asset-operator matching.
\end{enumerate*}
The effectiveness of the proposed system has been validated through indoor and outdoor experiments in a construction setting for identifying the operator of a power tool. A physical prototype was developed to evaluate the algorithms in a realistic setup. The results demonstrated a median accuracy of \qty{0.49}{\meter} in estimating the distance between assets and users, and up to 98.6\% in correctly matching users with their assets. 
\end{abstract}

\begin{IEEEkeywords}
iot, edge computing, sensor network, signal processing, cloud computation, embedded systems, low-power, bluetooth low energy, tracking
\end{IEEEkeywords}
\acresetall

\section{Introduction}
In recent years, the proliferation of connected devices has not only changed everyday life, but has also transformed the industrial sector~\cite{bajic21_indus, rikalovic22_indus}. At the center of this is the broad availability of \ac{IoT}, which is establishing into smart homes, smart cities, and smart manufacturing, among others~\cite{zanella14_inter_thing_smart_cities, misra22_iot_big_data_artif_intel}. Nevertheless, the construction industry still lags behind other industries regarding digitalization and efficiency~\cite{turner21_utiliz_indus, demirkesen21_inves_major_chall_indus}. Due to their inherent complexity involving different activities, stakeholders, and dynamic environments, construction projects are prone to risks such as delays, cost overruns, quality defects, and accidents~\cite{alvand21_ident_asses_risk_const_projec}. 

Asset-tracking solutions have gained prominence in industrial contexts to tackle some of these issues, as evidenced by their successful commercialization~\cite{huffstutler_23_hilti_adds_more_tool}. Yet, one significant gap is the lack of precise tracking of which operator uses which asset~\cite{sherafat20_autom_method_activ_recog_const_worker_equip}.
Addressing this gap provides valuable insights that can be exploited to improve efficiency and safety across various industries. Examples include:
\begin{itemize}
  \item \textbf{Healthcare Settings}: Tracking which medical professional uses a specific device (e.g., mobile X-ray or ultrasound machines), improves patient safety by ensuring only authorized personnel operate critical equipment, and facilitates timely maintenance~\cite{fatima22_patien_safet,luschi24_ohio}.
  \item \textbf{Manufacturing Environments}: Monitoring operators and machines (e.g., welding equipment, forklifts) enhances maintenance scheduling and detects the need for retraining, reducing breakdowns and downtime, while optimizing operational costs~\cite{zhao20_iot_edge_comput_enabl_collab}.
  \item \textbf{Construction Projects}: Operator-to-asset tracking on construction sites offers insights into tool usage, productivity, and equipment management. This helps to identify misuse, reduce wear and tear, and ensure compliance with safety regulations~\cite[p.~195]{edirisinghe18_digit_skin_const_site}. 
\end{itemize}
By knowing the right operator of a specific asset and the duration of the performed activity, it is possible to automatically generate asset-usage reports, check operators' tools safety training, and open a pipeline to allow easier digitalization of job sites. In turn, this additional granularity can further improve the existing asset management approaches~\cite{khurshid23_in_depth_survey_demys_inter, ranjithapriya20_study_factor_affec_equip_manag}.

In the pursuit of achieving asset-operator matching, several technical approaches can be explored. Among these, biometric systems integrated into assets, such as fingerprint readers on power tools, could be considered. Additionally, complex \ac{RTLS} deployed within job sites offer precise tracking capabilities for both users and assets~\cite{li16_real_time_locat_system_applic_const}. However, it is essential to acknowledge that these approaches face practical challenges when applied in dynamic industrial environments, particularly within the construction sector. The inherent complexities, high power consumption, and associated costs render them less feasible for widespread adoption. 

In the context of this paper, a novel approach for asset-operator matching is introduced. The fundamental premise underlying this approach is that the operator of a handheld asset must be the one nearest to the asset when it is active. Therefore, to pursue this idea, the following have to be monitored:
\begin{itemize}
    \item the distance between users and active assets, and 
    \item the real-time status of the asset. Specifically, whether it is active, e.g., in the case of a power tool, if its motor is running. This distinction is crucial for focusing the analysis on only the active assets.  
\end{itemize}

In implementing this novel approach, it is crucial to recognize that pinpointing the exact absolute distance between a user and an asset is not the primary objective. The essence of the method lies in accurately identifying the nearest operator to an active asset. This means that absolute accuracy in measuring the distance between the asset and the operator is not necessary. What is important is the relative distances between operators and the asset. For example, if Operator \(A\) is estimated to be at a distance of \qty{2}{\meter} and Operator \(B\) at \qty{4}{\meter} from asset \#1, despite their actual distances being \qty{1}{\meter} and \qty{3}{\meter} respectively, the system effectively discerns that Operator \(A\) is the nearest.

Given that most of the already deployed asset management systems are \ac{BLE}-based (e.g., Hilti On!Track) and that \ac{BLE} \ac{RSSI} is a function of the distance, the proximity of the user to the asset can be estimated. However, raw \ac{RSSI} values are affected by various phenomena, e.g., multi-path effect, leading to inaccurate measurements. To address this challenge, Bayesian filters can be employed. These filters leverage a physical model to exploit all available information and yield a more precise distance estimation~\cite{mackey20_improv_ble_beacon_proxim_estim}. 

Moreover, asset management \ac{IoT} devices are generally equipped with at least an accelerometer~\cite{giordano21_smart, krishnan21_real_time_asset_track_smart_manuf, magno19_smart, murphy15_wsn}. This capability can be utilized to monitor the activity status of the asset, allowing a focus solely on active assets. In our use-case it is assumed that no more than 15 active assets and operators are within \ac{BLE} range to each other. 
Leveraging these insights, the contributions of this paper are the following:
\begin{itemize}
    \item Novel approach: the proposed method is based on the insight that the operator of a handheld asset must be nearest to the asset when it is active. Each operator wears a wearable device that supports \ac{BLE} and \ac{LTE-M} connectivity. The asset is retrofitted with a \ac{BLE}-based tracking device equipped with an accelerometer. This device periodically broadcasts a message containing information about the asset's activity status estimated with a TinyML algorithm. The wearable receives this message and, if the asset is active, calculates the distance. All user-asset distances are sent to the cloud, where another algorithm sorts and estimates the nearest user to the active asset, assigning them as the asset's user.
    \item Accurate and energy efficient asset-user distance estimation at the edge: The proposed approach optimizes power consumption across all components. To achieve this, the study explores the possibility to enhance the accuracy of distance estimations from \ac{RSSI} measurements applying Bayesian filters on ultra-low power asset management \ac{IoT} devices and wearable devices.
    \item Real-world validation: The algorithms have been ported on previous developed hardware prototypes, i.e., \textsc{SmartTag}~\cite{giordano21_smart}  and \textsc{EcoTrack}~\cite{giordano23_energ_aware_adapt_sampl_self}, and tested in a realistic setup. The system was validated through extensive indoor and outdoor experiments in real construction settings, demonstrating its practical applicability. 
    
\end{itemize}

The rest of the paper is organized as follows: \cref{sec:related-works} reviews the state-of-the-art in asset-tracking and wireless distance estimation. \cref{sec:system-architecture} presents the high-level system architecture, and \cref{sec:algorithms} gives an in-depth view of the applied algorithms. \cref{sec:experimental-setup} describes the experimental setup and the experiments, while \cref{sec:results} reports and discusses the results. Finally, \cref{sec:conclusion} concludes the paper and outlines future work.
\section{Related Works}\label{sec:related-works}
Tracking assets in industrial environments, including construction sites, is nothing new. As early as 2005, Goodrum et al.~\cite{goodrum06_applic_activ_radio_frequen_ident} presented a \ac{RFID}-based system in which active \ac{RFID} tags are integrated into power tools to track them. The aim was to find lost tools. In an experimental evaluation, the tools could be found with a \ac{PDA} up to a distance of \qty{10.5}{\meter}, depending on the environment. 
A similar approach was taken by Goedert et al.~\cite{goedert09_autom_tool_track_const_site}, whereby utilization tracking was carried out by recognizing which power tools were still in the depot. An \ac{RFID}-equipped cabinet was able to determine with an accuracy of up to 98\% whether a tool is lying in the cabinet or is currently in use.
Kwon et al.~\cite{kwon21_uwb_mems_imu_integ_posit} developed a \ac{UWB}-based \ac{RTLS} in 2021 to track tools across the entire construction site with an accuracy of up to \qty{13.3}{\milli\meter}.

The systems presented above are either limited in terms of only detecting whether an asset is currently in use or not, or requiring a widespread anchor setup to ensure accurate localization across the entire industrial environment. The goal of this paper is different: we try to find out when an asset is used, and in particular by which operator, without an \ac{RTLS} infrastructure, such that the setup effort and costs are minimized. In order to achieve such a system, the first step is to recognize the distance between users and assets with simple \ac{IoT} devices.

The most common methods for distance measurement using radio frequency are based on \ac{RFID}, \ac{RSSI}, \ac{ToF}, or \ac{MCPD}~\cite{zafari19_survey_indoor_local_system_techn}. While \ac{RFID} is limited in range~\cite{goodrum06_applic_activ_radio_frequen_ident}, \ac{ToF} requires a high bandwidth to detect precisely \ac{LOS}, requiring a dedicated transmission technology like \ac{UWB}. \ac{MCPD}-based ranging requires the bidirectional exchange of multiple messages, thus consuming significantly more power when compared to \ac{RSSI}-based approaches~\cite{flueratoru20_energ_consum_rangin_accur_ultra, koenig11_multip_ieee, cortesi23_compar_between_rssi_mcpd_based}.

\ac{RSSI}-based approaches are based on the correlation between the received signal strength and the transmission distance~\cite{friis46_note_simpl_trans_formul}. The decrease of the signal strength can be modeled as logarithmic function of the distance, as shown in \cref{eq:rssi}~\cite{pu18_indoor_posit_system_based_ble}:
\begin{equation}\label{eq:rssi}
    \mathit{RSSI}(x)=\mathit{RSSI}(x_0)-10n\cdot\log_{10}\left(\frac{x}{x_0}\right)
\end{equation}

\ac{RSSI}-based distance measurements are widely used due to the availability in most radio frontends~\cite{balakrishnan21_rssi_based_local_track_spatial}. Furthermore, they require only unidirectional transmission, which reduces power consumption at the transmitter end. The widespread use of \ac{BLE} \acp{SoC} in modern smartwatches, smartphones, and more makes them an ideal candidate for various \ac{IoT} applications, including localization. During the COVID pandemic, it was widely utilized to monitor social distance and to keep track of spreading paths by exploiting \ac{RSSI} measurements\cite{castiglione21_role_inter_thing_to_contr, pathak21_iot_to_rescue, kumar21_social_distan_bluet_low_energ} -- a use case very similar to ours.

Although the theoretical equation suggests a strong relationship between \ac{RSSI} and distance, it does not account for all physical aspects in practice.
Due to the following factors:
\begin{enumerate*}[label=\textit{\roman*)}]
\item inaccurate measurements~\cite{agarwal21_deepb},
\item improper antenna matching~\cite{schulten19_crucial_impac_anten_diver_ble},
\item multipath fading and reflections~\cite{he22_tackl_multip_biased_train_data},
\item antenna directivity~\cite{schulten19_crucial_impac_anten_diver_ble}, and
\item attenuations due to obstructed line of sight~\cite{naghdi19_trilat_with_ble_rssi_accoun}
\end{enumerate*}, the relationship between distance and \ac{RSSI} is highly complex. As the distance increases, the impact of these errors becomes more significant, resulting in lower accuracy compared to methods based on \ac{ToF} or \ac{MCPD}. Consequently, the distance from a received signal power can only be predicted to a limited extent.

More accurate modeling of the path-loss that accounts for all factors in different industrial environments is complex, with no single model suitable for universal use~\cite[p.~27ff]{goldsmith05_wireless_communication}. Despite the many sources of errors, various algorithms have been explored to optimize the accuracy of range measurements based on \ac{RSSI} using \ac{BLE}~\cite{mackey20_improv_ble_beacon_proxim_estim, lee16_method_improv_indoor_posit_accur, qathrady17_improv_ble_distan_estim_class, gomez19_monit_harnes}, providing more reliable distance estimation.

\begin{table*}[htpb!]
    \centering
    \caption{Comparison of state-of-the-art approaches to improve \acs{BLE}-based \acs{RSSI} localization}
    \label{tab:comparison_distanceestimation}
    \renewcommand{\arraystretch}{1.3}
    \begin{tabular}{@{}lllllll@{}}
    \toprule
         &  IOT'20~\cite{mackey20_improv_ble_beacon_proxim_estim} & IPIN'23~\cite{debnath23_proxim_estim_ble_rssi_uwb}&MIS'16~\cite{lee16_method_improv_indoor_posit_accur} & MSWIM'17~\cite{qathrady17_improv_ble_distan_estim_class} & MEAS'19~\cite{gomez19_monit_harnes} & \textbf{This work}\\
    \midrule
    Algorithm & \acs{KF}, \acs{NI}, \acs{PF} & \acs{KF}+\acs{RF} & \acs{EKF} & diff. \acs{ML} & \acs{EKF} & \acs{EKF}\\
    Measurement scheme & static & static & static & static & \textbf{dynamic} & \textbf{dynamic}\\
    Localization setup & ranging & ranging & 6 anchors & ranging & ranging & ranging\\
    Processing unit & smartphone & \acs{PC} & server & \acs{PC} & ESP32 & nRF52833\\
    Measured range & \qty{0.5}{\meter} -- \qty{3.0}{\meter} & \qty{1.0}{\meter} -- \qty{8.0}{\meter} & \qtyproduct{12x11}{\meter} & \textbf{\qty{0.5}{\meter}--\qty{22}{\meter}} & \qty{0.0}{\meter}--\qty{7.0}{\meter} & \qty{0.0}{\meter}--\qty{6.0}{\meter}\\
    Accuracy & \textbf{\qty{0.27}{\meter} (\acs{PF})} & \qty{0.31}{\meter} & \qtyproduct{0.26x0.28}{\meter} & \qty{0.50}{\meter} & \(\approx\qty{1.00}{\meter}\) & \qty{0.49}{\meter}\\
    Deployment effort & medium & medium & high & medium & \textbf{low} & \textbf{low}\\
    \bottomrule
    \end{tabular}
\end{table*}

Mackey et al.~\cite{mackey20_improv_ble_beacon_proxim_estim} developed a mobile application that uses three Bayesian filtering techniques to process the \ac{RSSI} measurements received from \ac{BLE} beacons and estimate the distance between the beacon and the receiver. They focused on the following Bayesian filters:
\begin{enumerate*}[label=\textit{\roman*)}]
    \item a \ac{KF}, which assumes a linear model with white and Gaussian noise; 
    \item a \ac{PF}, which uses a set of consecutive samples to represent the posterior distribution; and 
    \item a \ac{NI} filter, which uses a kernel density estimator to approximate the posterior distribution.
\end{enumerate*}
The paper reported that with Bayesian filters, the distance estimation accuracy can be improved by up to 39\% compared to the moving average when the beacon and the receiver are within \qty{3}{\meter}. In a large room, the \ac{PF} showed the highest improvement of 39\% on average, the \ac{KF} showed an improvement of 25.6\%, and the \ac{NI} by 23.25\%. In a small room, the performance of the filters was closer together, with the \ac{PF} improving by 31\% on average, the \ac{NI} by 30\%, and the \ac{KF} by 28\%. Finally, the authors attribute the marginal superiority of the \ac{PF} over the \ac{KF} to its non-parametric approximation of the system, which fits the path-loss model better. Furthermore, they state that the runtime of the beacons is up to 21.4 months (Estimote), without providing the exact size of the battery.  Due to the higher computational costs of the \ac{PF}~\cite{stelzer17_compar_partic_filter_exten_kalman} and \ac{NI} filter compared to the \ac{KF}, while there is only a marginal difference between the filters, especially in small spaces, our system is based on the \ac{EKF}.
\par
Debnath et al.~\cite{debnath23_proxim_estim_ble_rssi_uwb} improved the \ac{RSSI}-based ranging using a \ac{KF} followed by an \ac{RF} model with 100 decision trees. The measurements of each \ac{BLE} advertisement channel are thereby processed individually before the results are combined by averaging. The experimental evaluation at static measurement positions from \qty{0.5}{\meter} to \qty{8.0}{\meter} showed an average accuracy of \qty{0.31}{\meter}. The computations are performed on a \ac{PC}.
\par
Lee et al.~\cite{lee16_method_improv_indoor_posit_accur} proposed an \ac{EKF} for enhancing the deteriorated \ac{RSSI}-based position estimation caused by noise, motion, and fading. In contrast to our approach, fixed \ac{BLE} anchors and a two-dimensional motion model are used, and the computation is performed on a server. The \ac{EKF} handles the nonlinearities and uncertainties of both the motion and measurement models. Due to its filter characteristics, it provides a more accurate estimate of the smartphone's position than trilateration using raw \ac{RSSI} signals. Experimental measurements have been conducted in an indoor environment using static measurement positions. Their method achieved a localization accuracy of \qty{0.26}{\meter} and \qty{0.28}{\meter} in x- and y-coordinates over 100 measurements in a \qtyproduct{12x11}{\meter} room.
\par
In order to further increase the accuracy of distance estimation based on \ac{RSSI}, Al Qathrady and Helmy~\cite{qathrady17_improv_ble_distan_estim_class} integrated the \ac{RSSI} and knowledge of the \ac{BLE} TX power level into several parametric and \ac{ML} models running on a connected \ac{PC}. The study was conducted based on 1.8 million indoor recordings from static positions at distances between \qty{0.5}{\meter} and \qty{22}{\meter}. With the integration of the TX power level, they reduced the mean average estimation error by up to 46\% compared to their model without TX power level integration. The final precision of the distance estimation was \qty{0.5}{\meter}.
\par
Gómez-de-Gabriel et al.~\cite{gomez19_monit_harnes} deployed \ac{BLE} beacons in risky areas where a lifeline connected to a harness is needed to monitor the proper usage of harnesses on construction sites. Having an additional beacon on the lifeline itself and a receiver on the harness, they try to estimate whether workers are connected to the lifeline or not in risky areas. Their distance estimation approach is based on an \ac{EKF} filter running on an \textsc{ESP32} \ac{MCU}. The system state model used for the \ac{EKF} is a \textit{no-motion} motion model~\cite{madrigal12_simult_local} where state transitions (i.e., worker movement) are modeled as Gaussian stochastic perturbations. The model is based on the assumption of a maximum worker speed of \qty{0.5}{\meter\per\second} with a probability of 95\%. Their evaluation in a dynamic setup shows a distance estimation accuracy of \(\approx\qty{1}{\meter}\).

Our approach results in slightly worse accuracies when compared to more sophisticated algorithms such as \ac{NI} or \ac{RF}, as shown in \cref{tab:comparison_distanceestimation}. However, due to the limited computational and energy resources and the requirement to run in real-time, this represents the trade-off taken to perform the calculation on an \ac{MCU} instead of a computer or server. Furthermore, our measurement is in a dynamic setup and not at predefined static distances.
\section{System Architecture}\label{sec:system-architecture}
\begin{figure}[htpb!]
    \centering
    \includesvg[width=\columnwidth]{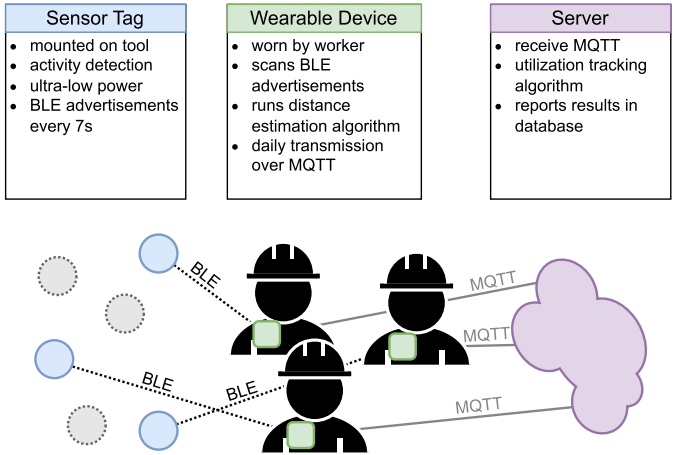}
    \vspace*{-5mm}
    \caption{Overview of the three parts of the system: sensor tags mounted on inactive assets (gray), sensor tags mounted on active assets (blue), wearable devices close to the active sensor tags (green) and a centralized server (violet).}
    \label{fig:system-overview}
\end{figure}

The proposed system consists of three parts (\cref{fig:system-overview}): A sensor tag, a wearable device (worn as a badge), and a centralized server. The sensor tag is a smart \ac{BLE} node with an activity detection feature used to identify the assets. It advertises the asset's current status every seven seconds, reporting whether it is being used or not. The key factor here is that the tag is characterized by ultra-low power consumption, ensuring a lifetime of multiple years in battery-powered mode when it is retrofitted onto existing tools. 

The wearable device, carried by each user and supporting \ac{BLE} and cellular connectivity, scans for \ac{BLE} advertisements of such a sensor tag and records its \ac{RSSI}. In the proposed system architecture, the wearable operates as an edge device, pre-processing the real-time distance measurements to the individual tags and transmitting the aggregated data to a server.

The resulting high variance of the \ac{RSSI} values can be filtered to remove noise and minimize the deviation from the theoretical path-loss equation proposed by~\cite{mackey20_improv_ble_beacon_proxim_estim, lee16_method_improv_indoor_posit_accur, qathrady17_improv_ble_distan_estim_class, gomez19_monit_harnes} and discussed in \cref{sec:related-works}.

Two strategies can be considered for the distance estimation: the first strategy involves transmitting all raw data to the server, where \ac{ML}-enhanced processing techniques such as \ac{DL}, \ac{NI}, or \ac{RF} could provide a more  accurate distance estimation. However, this approach requires a considerable amount of data to be transferred from the wearable device to the server. 

This high data transmission demand would significantly limit the lifetime of the battery-powered device due to the high power consumption required for data transmission (as shown in ~\cref{tab:power_profile}), and would also raise transmission  costs.

The second option is to exploit the wearables' computing capabilities for estimating the distance between assets and users. As a result, the transmission size can be significantly restricted and kept constant, regardless of the session duration of the asset usage. 

To fully benefit from the energy savings of the reduced amount of transmitted data, the computational expenses for distance calculation on the wearable device must be kept low. To achieve this, a 1-state \ac{EKF} algorithm was selected in this work, which offers good performance as demonstrated in previous studies~\cite{lee16_method_improv_indoor_posit_accur, gomez19_monit_harnes}, and requires low computational effort.

Finally, using its cellular interface, the wearable device transmits the estimated distances for each active period of an asset to a centralized server once per day.

After the new distance data arrives from the wearable devices, it is written to a database by the server. The data are sorted by time and organized in sessions. Each of these sessions groups an active period of an asset with the distance data of the users. In solving an optimization problem, the relation between an active asset and its correct user is estimated. The results are then made available to the end user in a dashboard.
\section{Algorithms}\label{sec:algorithms}
The main challenge of the algorithm running on the wearable devices lies in achieving an accurate real-time distance estimation between the sensor tags attached to individual assets and the wearable badge. Due to these devices' limited computation and energy resources, computationally lightweight and energy-efficient algorithms are needed. Furthermore, the distance estimations must be accurate in the face of environmental noise and measurement uncertainties inherently given by the \ac{RSSI} recordings.

Simultaneously, the asset-user tracking algorithm, which operates on distance measurements aggregated from multiple wearables on a central server, faces its own challenges. While it benefits from greater computational resources compared to wearable devices, it must effectively handle the dynamic nature of tool usage scenarios, where multiple users may interact with each other and the various assets concurrently.

The proposed system to assign assets to their users consists, therefore, of two algorithms:
\begin{itemize}
    \item A distance estimation algorithm optimized and evaluated on the low-power wearable device.
    \item An asset-user tracking algorithm, based on an optimization problem, evaluating the distance measurements of all badges and assigning the most probable matching of tags and badges together with a probability classifier, yielding the trust level 'SURE' or 'UNSURE'.
\end{itemize}
\subsection{Distance Estimation}\label{sec:embedded-algo}
    The state variable of the dynamic system that is attempted to be estimated is the distance \(x\) between the tag and the badge. The model is based on the assumption that the distance between the user and the used assets is virtually constant, with random variations around this distance. Specifically, the relative movement between assets and operators can be modeled as random noise; therefore, a \textit{no-motion} motion model~\cite[p.~169ff]{madrigal12_simult_local} was used. In this model, the state evolution of the kinematic equation consists of propagating the state (i.e. the distance) from the last step without changes; the movement between the operators and assets is modeled as a zero-mean Gaussian stochastic perturbation \(w_k\) as in \cref{eq:q_k}:
    \begin{align}\label{eq:non-motion-model}
        x_k &= x_{k-1} + w_k, & w_k & \sim \mathcal{N}(0,Q_k)
    \end{align}
    where the process noise covariance \(Q_k\) is chosen as
    \begin{align}\label{eq:q_k}
        Q_k &= \frac{v_{max}^2}{\mathcal{X}_{1,c}^2} = \qty{0.1275}{\meter\squared\per\second\squared}
    \end{align}
    with \(v_{max}\) the expected maximum relative velocity of the operators with respect to the assets (\qty{0.7}{\meter\per\second}) in 95\% of all times (\(c=0.05\)) and \(\mathcal{X}_{1,c}^2\) the support value of a Chi-squared distribution marking the beginning of an upper tail with area \(c\) and 1 degree of freedom.
    \par
    The observation model is defined in \cref{eq:observation} and follows the path-loss model, with the observation variable \(z\) representing the \ac{RSSI} measurement.
    \begin{align}\label{eq:observation}
        z_k &= h(x_k) + v_k, \quad v_k \sim \mathcal{N}(0,R_k)\\
        h(x) &= \mathit{RSSI}(x_0)-10n\cdot\log_{10}\left(\frac{x}{x_0}\right)
    \end{align}
    The measurement noise covariance \(R_k\) in \cref{{eq:r_k}} was evaluated experimentally using all measurement values in \cref{fig:rssi-vs-d}.
    \begin{equation}\label{eq:r_k}
        R_k=\sigma^2=43.53
    \end{equation}

    \par
    The \ac{EKF} is then composed of two steps, the prediction and the update step, as in~\cite[p.~310ff]{grewal08_kalman_filtering}, with
        \begin{equation*}
            H_k=\left.\frac{\partial h}{\partial x}\right\vert_{\hat{x}_{k|k-1}}=\frac{-10n}{\log(10)\cdot \hat{x}_{k|k-1}}
        \end{equation*}
    \par

    Due to the high variance of the \ac{RSSI} measurements and the non-linearity of the measurement model, the choice of the initial value can severely impact the filter's evolution and is, therefore, crucial~\cite{reif99_stoch_stabil_discr_time_exten_kalman_filter}. 
    
    The initial state estimation was, therefore, based on the path-loss model but restricted to realistic distances:
    \begin{equation}
        \hat{x}_{0|0} = \max{\left(\min{\left(h^{-1}(\mathit{RSSI}),\qty{20}{\meter}\right)},\qty{0.5}{\meter}\right)}
    \end{equation}
    Therefore, if the initialization is performed using information from an \ac{RSSI}-outlier (\ac{RSSI} below \qty{-58.75}{\decibel} or above \qty{-42.56}{\decibel}), as determined from the interpolated path-loss curve in~\cref{fig:rssi-vs-d}, the initial distance is constrained to \qty{0.5}{\meter} and \qty{20}{\meter}, respectively. Outliers can result in an initial distance estimation of several hundred meters due to the logarithmic decay in path loss, despite the operator being very close (\qty{0.5}{\meter}) - or can lead to initial distances much closer than \qty{0.5}{\meter}, although being much further away. These distances are assumed to be realistic within our setup and are tuning parameters, as no operator is expected to be closer than \qty{0.5}{\meter} to a power tool when they receive the first advertisement, and no one is expected to be further than \qty{20}{\meter} from a tool when it is activated. Utilizing such outliers as the initial estimation can cause convergence of the filter to wrong estimates, due to the non-linearity of the measurement model~\cite{boutayeb97_conver_analy_exten_kalman_filter}. Hence, a maximal initial distance of \qty{20}{\meter} is assumed; the filter can subsequently adjust to greater distances if necessary, or converge to a closer distance if the operator is closer.
    \par
    Recapped, whenever a tag detects its asset to be active, it starts advertising its state in a \qty{7}{\second} interval. The wearable, constantly scanning, detects the advertisements and starts a new instance of the embedded \ac{EKF} algorithm based on the received \ac{RSSI} values from the advertisements. Once the advertisements stop (i.e. the tag went back to inactive), the final distance estimate is transmitted to the server together with the start and stop time.

\subsection{Asset-User-Tracking}\label{sec:cloud-algo}
    As the wearables do not have a global view of all wearables-to-tags distances, a centralized asset-user tracking algorithm has been implemented to match any used asset with its corresponding user (i.e., the wearable).

    After collecting distance estimations of all wearables with respect to the tags, the basic idea is to assign a user to each active period of an asset. At the beginning, the server has a list of all active periods of the sensor tags sorted by their start times. A single device can have multiple active periods and thus be present multiple times in this list. The algorithm starts with the tag that became active first (i.e., the first in the list). The user, respectively wearable device, nearest to the tag is assigned to this active period of the asset. Subsequently, the nearest user is then assigned to each additional active period of a device, provided that the user is not already using an asset. This problem can be formulated as the following optimization problem:

    Given is a set of all users and assets (can be extracted from the information transmitted by the wearable devices):
    \begin{align}
    \begin{split}
        I&: \text{set of all wearable devices (users) } i\\
        J&: \text{set of all sensor tags (assets) } j\\
        t&: \text{current time step}\\
        d(i,j,t)&: \text{estimated distance between }i\text{ and }j\text{ at time }t\\
        b_{jt}&: \text{state of the asset }j\text{ at time }t
    \end{split}
    \end{align}
    with \(b_{jt}\) being 1 if the asset is active at time \(t\) and 0 otherwise.
    
    The decision variables are the assignment of a user \(i\) to an asset \(j\), namely \(a_{ijt}\), as well as a trust level \(c_{ijt}\) belonging to this assignment:
    \begin{align}
    \begin{split}
        a_{ijt} &= \begin{cases}
            1 & \iff i\text{ is using }j\text{ at time }t\\
            0 & \text{otherwise}
        \end{cases}\\
        c_{ijt} &= \begin{cases}
            1 & \iff \text{the assignment } a_{ijt} \text{ is likely correct}\\
            0 & \text{otherwise}
        \end{cases}
    \end{split}
    \end{align}
    Therefore, the optimization goal is to assign the nearest operator-asset combination at each time step, and therefore providing the assignments \(a_{ijt}\) and the trust level classifier \(c_{ijt}\).
    \begin{align}
    \text{minimize}\hspace{0.5cm} & \sum_i\sum_j\sum_t a_{ijt}\cdot d(i,j,t)\label{eq:optimization}\\
    \text{subject to}\hspace{0.5cm} & \sum_i a_{ijt} = \begin{cases}
            1 & \iff b_{jt}=1,\ \forall{j,t}\\
            0 & \iff b_{jt}=0,\ \forall{j,t}
        \end{cases}\label{eq:singleuser}\\
        & a_{ij(t+1)}=a_{ijt}\ \iff b_{j(t+1)}=b_{jt}\label{eq:continuity} \\
    \begin{split}\label{eq:sure}
        & c_{ijt} = 1 \ \iff (a_{ijt}=1)\ \land\\
        &\quad (\lvert d(i,j,t) - d(k,j,t) \rvert > \qty{0.75}{\meter})\ \forall k\in I\
    \end{split}
    \end{align}
    The optimization is subject to the constraint that each asset can only be used by exactly one user at any given time and that the asset must be active (\cref{eq:singleuser}). Finally, the system is restricted to the extent that the operator cannot change during a continuous active time (session) of a device (\cref{eq:continuity}). Assignments \(a_{ijt}\) are provided with a trust level classifier, namely \(c_{ijt}\), to account for residual \ac{EKF} distance estimation errors. Whenever the estimated distance between the matched user and the next nearest user is less than \qty{0.75}{\meter} (as described in~\cref{eq:sure} with \(\land\) being the logical conjunction operation), a classifier of 'UNSURE' is assigned to this matching. A classifier of 'SURE' means that the solution of the optimization problem matches the actual asset-user pairing with a high degree of certainty.

    To solve the problem,~\cref{alg:asset-op-matching} is executed for each time step, starting from \(t=0\). In an initial step, inactive tools are unassigned (\(\sum_i a_{ijt} = 0\), \cref{eq:singleuser}) and the continuity constraint~\cref{eq:continuity} is applied (\cref{alg:asset-op-matching:cleared,empty1,empty2,alg:asset-op-matching:continuity}) in order to reduce the search space. 
    Then, an exhaustive search is executed as in~\cref{alg:asset-op-matching:search}. 
    
    The exhaustive search iterates through the reduced sets \(I\) and \(J\), evaluating the cost function of~\cref{eq:optimization} and returning the assignment leading to the lowest cost value. In most cases, only a single asset is started or stopped simultaneously during a given time step, reducing \(J\) to a singleton, and the search to finding the nearest operator of the asset.

    \begin{algorithm*}
        \caption{Iterative Asset-Operator Matching to find \(a_{ijt}\)}\label{alg:asset-op-matching}
        \begin{algorithmic}[1]
        \Function{Matching}{$I,\ J,\ t,\ d,\ a_{0:t-1},\ b_{0:t}$}
            \State $a_{ijt} \gets 0\ \forall i\in I,\,j\in J$ \Comment{Initialize solution for time step \(t\) with zero}
            \ForAll{$j\in J$} 
                \If{$b_{it}=0$}\Comment{Inactive assets are not assigned to an operator.}\label{alg:asset-op-matching:cleared}
                    \State $a_{ijt}\gets 0\ \forall i\in I$\label{empty1}
                \ElsIf{$b_{it}=b_{i(t-1)}=1$}\Comment{Assets active in previous time step do not change operator.}\label{empty2}
                    \State $a_{ijt}\gets a_{ij(t-1)}\ \forall i\in I$\label{alg:asset-op-matching:continuity}
                \Else\Comment{Assign newly activated assets to the correct operator}
                    \State exhaustiveSearch() \label{alg:asset-op-matching:search}\Comment{by searching through the solution space}
                \EndIf
            \EndFor
        \EndFunction
        \end{algorithmic}
    \end{algorithm*}

    For the practical implementation of the system (\cref{{sec:system-architecture}}), the following limitations apply:
    \begin{itemize}
    \item To distinguish between two subsequent asset activities of an asset, there must be a minimum pause of \qty{21}{\second} to be considered as two activities (3 advertisements, \qty{7}{\second} each).
    \item With a classifier of 'UNSURE', there is an increased risk that the solution is incorrect, or multiple users have been working on the same tool. The distance of \qty{0.75}{\meter} comes from the assumption that two users do generally not stand closer to each other.
    \end{itemize}
\section{Experimental Setup}\label{sec:experimental-setup}
In order to evaluate the accuracy of the algorithms, an experimental setup was set up in the construction sector, where construction workers (our users) are to be matched to the power tools used (our assets). 
\subsection{Hardware setup}
    The \textsc{SmartTag} has been used as the sensor node for the experimental evaluation. It is a smart \ac{BLE} beacon targeted for ultra-low power consumption and built for easy retrofit on existing power tools.
    The \textsc{SmartTag} is built around an nRF52810 \ac{BLE} \ac{SoC} from Nordic Semiconductor. The \ac{SoC} integrates an ARM Cortex-M4F with \ac{BLE} 5.3 connectivity, \qty{192}{\kilo\byte} Flash, and \qty{24}{\kilo\byte} \acs{RAM}. Evaluating the measurements of the ultra-low power accelerometer \textsc{IIS2DLPC} from \textsc{STMicroelectronics}, it advertises every \qty{7}{\second} the asset's current state: active working (\textit{Usage} class), any other non-active working phase present during construction activity (\textit{Transportation} class), and inactivity. A simple \ac{FFT}-based approach was used in the first implementation~\cite{giordano21_smart}, which only worked with one type of drill (Hilti TE 30-A36); then, a computational-heavy implementation using neural networks was implemented~\cite{giordano22_desig_perfor_evaluat_ultral_power}, followed by a \textsc{MiniRocket}-based version~\cite{giordano23_optim_iot_based_asset_utiliz_track}, which can classify activity on a variety of power tools (drills, hammers, etc.). The \textsc{MiniRocket}-based classification achieves an accuracy of 96.9\% across 16 tested power tools of different manufacturers. The \qty{7}{\second} interval was defined in~\cite{giordano23_optim_iot_based_asset_utiliz_track} as the optimum trade-off between update frequency and power consumption. Assuming a typical usage duration of power tools between \qty{30}{\second} and \qty{120}{\minute}~\cite{vergara08_hand_trans_vibrat_power_tools, edwards06_hand_vibrat_expos_from_const_tools}, this corresponds to at least four measurements of \ac{RSSI} values, which already provide enough informations for an initial distance estimation. Having a power consumption of only \qty{15}{\micro\watt}, its battery life is estimated to be more than three years on a \qty{675}{\milli\watt{}\hour} battery.

    \par

    As wearable device, the \textsc{EcoTrack}~\cite{giordano23_energ_aware_adapt_sampl_self} has been used. The wearable device is worn by each construction worker on the chest, which was found to be the most comfortable position during testing. However, the position can easily be adjusted if, for example, a smartwatch is used.
    The \textsc{EcoTrack} consists of an nRF52833, a \ac{BLE}-enabled Cortex-M4 from \textsc{Nordic Semiconductor}. It has been used to track the worker's condition using its \textsc{SARA R510M8S} \ac{GNSS} and \ac{LTE-M} module and an \textsc{LSM6DSLTR} \ac{IMU}. Achieving a power consumption of \qty{16}{\milli\watt} while being connected over \ac{LTE-M} and scanning for \ac{BLE} advertisement, it is prepared to operate for a multiple weeks on a single battery charge of \qty{10800}{\milli\watt{}\hour}. Using an encrypted \ac{MQTT} connection to the server, messages can be exchanged asynchronously by the \textsc{EcoTrack} and the server.

    \par

    The centralized server was implemented using AWS Lambdas, where a Lambda receives the data via MQTT and stores it in a database. New data then triggers a second Lambda, which solves the linear program from \cref{sec:cloud-algo} and makes the data available to the end user in a dashboard.

\subsection{Measurement setup}\label{sec:measurement-setup}
    To quantify the accuracy of the proposed system, tests were conducted both indoors and outdoors. During these tests, tags were mounted on various power tools (Hilti SFC 14-A, Hilti TE 30-A36, Hilti TE 706-AVR, Hilti SF 10W-A22), and a wearable was attached to the chest of each worker. During the indoor measurements, the exact position of all devices was recorded with the \textsc{OptiTrack} system\footnote{\url{https://optitrack.com/}, 08.02.2024}. The data obtained by the \textsc{OptiTrack} is used as ground truth, especially for distance estimation. During all measurements, a \textsc{Nikon Z6 II} video camera was used to record the test and as a sanity check for correct asset-user tracking. The following measurement tests were performed indoors:
    \begin{enumerate}
        \item One worker on a power tool, one person standing close by with a distance of \qty{0.5}{\meter}, \qty{1}{\meter}, \qty{2}{\meter}, \qty{3}{\meter} respectively.
        \item One worker on a power tool, two persons standing close by with distances of \qty{0.5}{\meter}, \qty{1}{\meter}, \qty{2}{\meter}, \qty{3}{\meter} each.
        \item Two workers working on a power tool each, working at distances of \qty{0.5}{\meter}, \qty{1}{\meter}, \qty{2}{\meter}, \qty{3}{\meter}.
        \item Two workers working on a power tool each, working at distances of \qty{0.5}{\meter}, \qty{1}{\meter}, \qty{2}{\meter}, \qty{3}{\meter}. One person standing close by with a distance of \qty{0.5}{\meter}, \qty{1}{\meter}, \qty{2}{\meter}, \qty{3}{\meter}.
        \item Three workers working on a power tool each, working at distances of \qty{0.5}{\meter}, \qty{1}{\meter}, \qty{2}{\meter}, \qty{3}{\meter}.
        \item Three workers working on a power tool each, working at distances of \qty{2}{\meter}. After \qty{2}{\minute} and \qty{4}{\minute}, they swap the power tools with each other. These results are presented in~\cref{fig:confusion-matrix}
    \end{enumerate}
    Each of those experiments was conducted for \qty{3}{\minute} to \qty{6}{\minute}. Data was collected with an advertising interval of \qty{0.5}{\second}; to replicate the 7-second interval, data was downsampled to match an advertisement interval of \qty{7}{\second}. In this way, a fourteen times larger 7-second dataset was obtained. The last experiment shows a dynamic use case, where workers change tools between them.
    \par
    In a later step, the following measurement tests were performed outdoors:
    \begin{enumerate}
        \item One worker on a power tool, one person standing close by with a distance of \qty{0.5}{\meter}, \qty{1}{\meter}, \qty{2}{\meter}, \qty{3}{\meter} respectively.
        \item Two workers on a power tool each, working at distances of \qty{0.5}{\meter}, \qty{1}{\meter}, \qty{2}{\meter}, \qty{3}{\meter}.
        \item Three workers on a power tool each, working at distances of \qty{0.5}{\meter}, \qty{1}{\meter}, \qty{2}{\meter}, \qty{3}{\meter}.
        \item Three workers on a power tool each, working at changing distances between each other.
    \end{enumerate}
    The last experiment shows a dynamic use case in the outdoor environment, where the workers change their working distances with respect to each other.
    \par
\begin{figure}[htpb!]
        \centering
        \includesvg[width=\columnwidth]{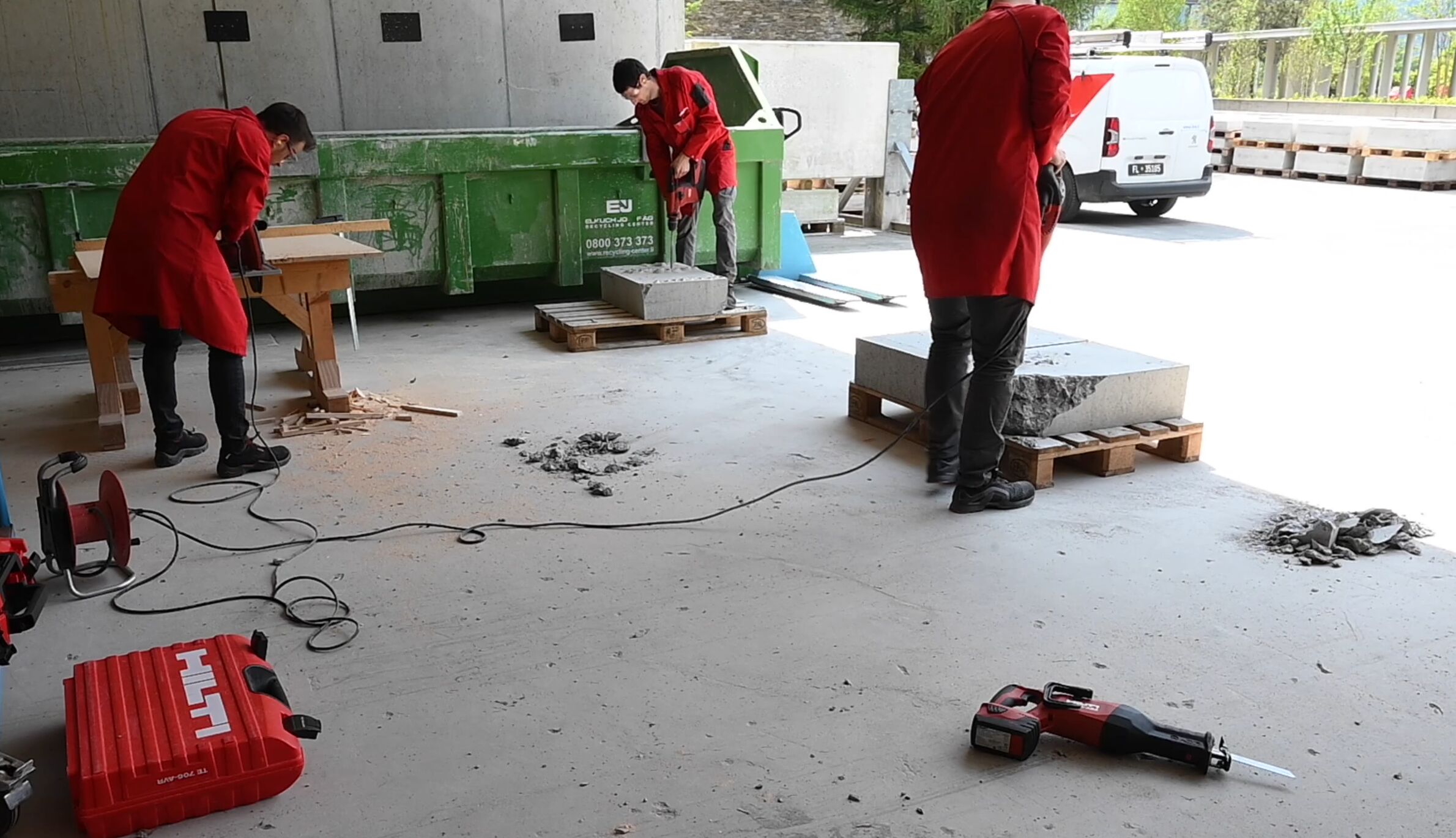}
    \vspace*{-5mm}
        \caption{Outdoor experimental setup. Three workers, each with their own asset, separated all by a distance \(d\) of \qty{2}{\meter}}
        \label{fig:outdoorsetup}
    \end{figure}

    \cref{fig:outdoorsetup} shows the outdoor experimental setup. The indoor setup was correspondingly the same.

    The power measurements were conducted using a \textsc{Keysight N6705C} with the \textsc{N6781A} module. The following measurements were taken:
    \begin{enumerate}
        \item Power consumption of the badge in scanning mode (i.e., only \ac{BLE} scanning)
        \item Power consumption of the badge during the calculation of the distance estimation algorithm (i.e., \ac{BLE} off, \ac{CPU} computing at \qty{64}{\mega\hertz})
        \item Power consumption of the badge transmitting values to the cloud (i.e., only \ac{LTE-M} active)
    \end{enumerate}
    The power consumption of the \textsc{SmartTag} has been reported and discussed in detail in~\cite{giordano23_optim_iot_based_asset_utiliz_track}.
\section{Results}\label{sec:results}
    This section presents the measurements performed to fit the path-loss equation to our \ac{RSSI} data. It determines the accuracy of the two algorithms explained in \cref{sec:embedded-algo} and \cref{sec:cloud-algo}. In particular, the accuracy of the calculated distances of the \ac{EKF} and \ac{CA} algorithms are compared to their ground truth. Afterwards, the results of the asset-to-user matching based on the aforementioned distance estimations are presented. Subsequently, each of the two implications is discussed, and the occurrence of the measurement errors is examined in more detail. Finally, the power consumption of the algorithm implemented on the wearable device is presented.

    \subsection{Path-loss fitting}
\begin{figure}[htpb!]
            \centering
            \includesvg[width=\columnwidth]{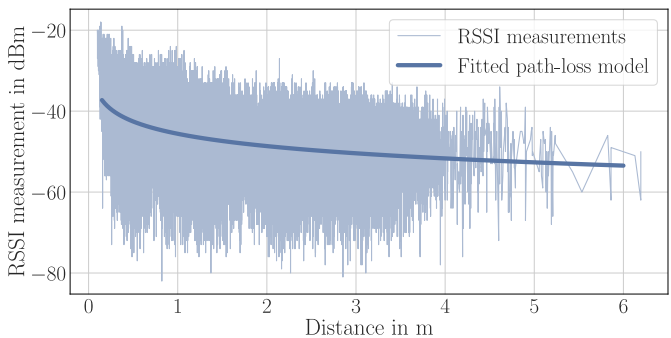}
    \vspace*{-5mm}
            \caption{\acs{RSSI} measurements versus ground truth distance together with fitted path-loss equation (\cref{eq:rssi}). Does not include results of dynamic measurements.}
            \label{fig:rssi-vs-d}
        \end{figure}

        \cref{fig:rssi-vs-d} shows \cref{eq:rssi} fitted to all our measurements (55'097 \ac{RSSI} values, \(n=1.011,\ x_0=\qty{1}{\meter},\ \mathit{RSSI}(x_0)=\qty{-45.6}{\decibel}\), experimental setup as in~\cref{sec:experimental-setup}). These values are used in the \ac{EKF} of the badge for estimating the distance between tag and badge.
        
        The variance of all raw \ac{RSSI} values with respect to the fitted path-loss curve is \qty{48.92}{\decibel\squared}, and the standard deviation is \qty{6.99}{\decibel}. It can be seen that individual \ac{RSSI} readings at a distance of \qty{6}{\meter} can also occur at a distance of less than \qty{50}{\centi\meter} - the map is, therefore, no longer bijective due to the aforementioned imperfections. However, the trend of a logarithmic curve can still be recognized. 
            
    \subsection{Distance estimation}
\begin{figure}[htpb!]
            \centering
            \includesvg[width=\columnwidth]{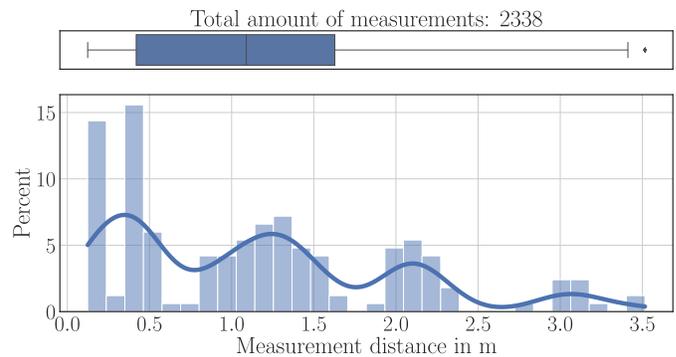}
    \vspace*{-5mm}
            \caption{Distribution of the 2338 performed measurements according to their ground truth distances. Does not include results of dynamic measurements.}
            \label{fig:measurement-distances}
        \end{figure}

        In our experiments' scope, a total of 2338 distance estimates were made at different measurement distances. These estimates are used in the following and compared to their ground truth distance to evaluate the accuracy and precision of the estimates. The amount of estimations per ground truth distance vary, as can be seen in \cref{fig:measurement-distances}. As the distance between a user and its asset during the experiments was always below \qty{0.5}{\meter}, 35.07\% of all estimations were carried in the range between \qty{0}{\meter} and \qty{0.5}{\meter}.
        To investigate the accuracy of the distance estimation, all distances estimated by the badges were compared with their ground truth values recorded by the \textsc{OptiTrack} system, and the estimation error was calculated. An overview of these results can be found in \cref{tab:distanzestimation}. 
        \begin{table}[htpb!]
            \centering
            \caption{Statistics of the distance estimation algorithm}
            \label{tab:distanzestimation}
            \renewcommand{\arraystretch}{1.3}
            \begin{tabular}{@{}ll@{}}
                \toprule
                Quantity & Measure\\
                \midrule
                2338 & Total amount of estimated distances\\
                \qty{0.49}{\meter} & Median Error\\
                \qty{1.18}{\meter} & Mean Absolute Error\\
                \qty{1.92}{\meter} & Root Mean Square Error\\
                \qty{1.63}{\meter} & Standard Deviation\\
                \qty{2.67}{\meter^2} & Variance\\
                \qty{11.78}{\meter} & Maximum (Absolute) Error\\
                \qty{0.00}{\meter} & Minimum (Absolute) Error\\
                \bottomrule
            \end{tabular}
        \end{table}

\begin{figure}[htpb!]
            \centering
            \includesvg[width=\columnwidth]{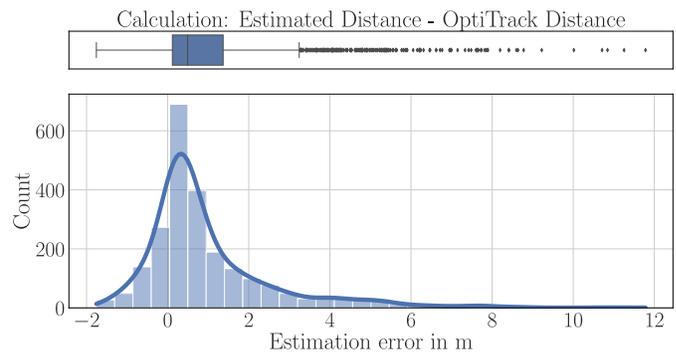}
    \vspace*{-5mm}
            \caption{Estimation error with respect to ground truth, evaluated over all 2338 estimations. Box-whisker plot highlighting the distribution of the values. Does not include results of dynamic measurements.}
            \label{fig:estimation-error}
        \end{figure}

\begin{figure*}
            \centering
            \includesvg[width=2\columnwidth]{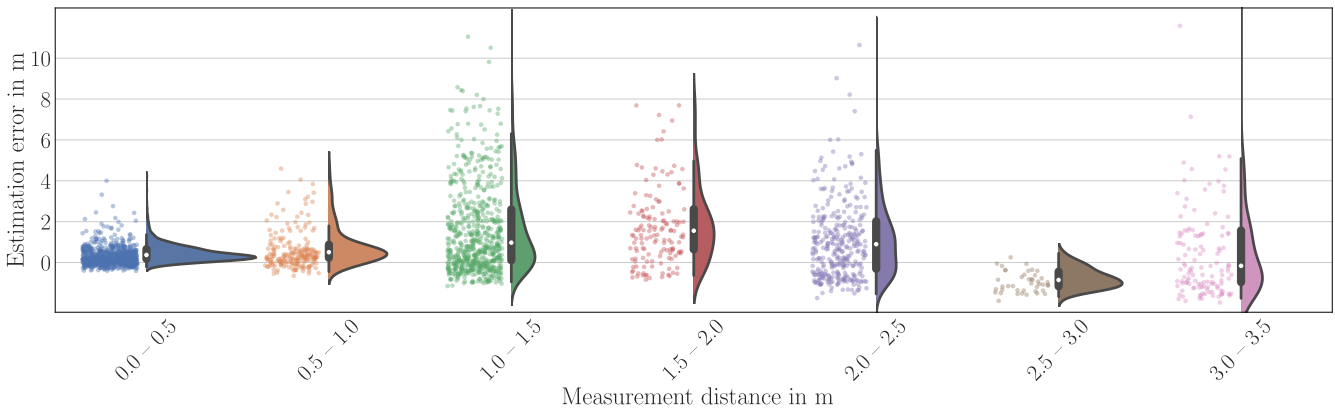}
            \caption{Distribution of estimation errors against their ground truth distances. The errors of each distance bin are represented as swarm and as half-violin with \ac{IQR} and Whiskers. Does not include results of dynamic measurements.}
            \label{fig:estimation-error-distribution}
        \end{figure*}

        The calculated estimation error is \qty{0.49}{\meter} in median with a standard deviation of \qty{1.63}{\meter} and a variance of \qty{2.67}{\meter^2}. These results, plotted as a histogram with associated \ac{IQR} ([\qty{0.12}{\meter},\ \qty{1.37}{\meter}]) and whiskers in \cref{fig:estimation-error}, show that 90.98\% of all estimations (represented by whiskers) are within [\qty{-1.76}{\meter},\ \qty{3.24}{\meter}].
        The breakdown of these estimation errors against their ground truth values, as shown in \cref{fig:estimation-error-distribution}, shows that the ground truth distance influences the estimation error: The median error for all ground truth distances smaller than \qty{1}{\meter} is \qty{0.37}{\meter}. The estimations especially lie close together, with the variance being \qty{0.34}{\meter\squared}. When the effective distance increases, the median slightly increases to \qty{0.86}{\meter} for all distances higher than \qty{1}{\meter}, and the variance grows to \qty{4.25}{\meter\squared}.
        \par
        This behavior can be explained by the fact that the logarithmic decay of the theoretical path-loss model (\cref{eq:rssi}) reacts much stronger to measurement inaccuracies and disturbances of the \ac{RSSI} the further the distance between the devices is. The distances between \qty{2.5}{\meter} and \qty{3}{\meter}  contributed to only 1.80\% of all estimations and were all recorded during a single experiment. They are, therefore, insufficient to draw statistical conclusions.

        If these results are compared with~\cite{gomez19_monit_harnes}, it is noticeable that the accuracy of our distance estimation is about \qty{0.5}{\meter} higher, even though the same \ac{EKF} is used. There may be several reasons for this, but mainly:
        \begin{itemize}
            \item A different path-loss model is used: while~\cite{gomez19_monit_harnes} uses a fit to the exponential function \(h(x)=a\mathrm{e}^{bx}+c\mathrm{e}^{dx}-o\), here the model as shown in \cref{eq:rssi} is used. Due to the smaller number of parameters, our model is less dependent on the environment, thus adapting more accurately to changes in the environment.
            \item The distance between the asset and the user is almost constant in our model, while the distance between the node and the tag in~\cite{gomez19_monit_harnes} changes between \qty{1}{\meter} and \qty{5}{\meter}. Due to the exponential decay of the path-loss, measurement errors are bigger on larger distances.
            \item Different radio frontends were used (~\cite{gomez19_monit_harnes}: \textsc{ESP32}), the measurement accuracy of \ac{RSSI} can therefore vary.
        \end{itemize}
    \subsection{Asset - User Matching}
        The 2338 distance estimations led to 980 matches between worker and tool. 546 of those matches are done in the indoor setup, and 434 in the outdoor setup.

        Following up on the trust classifier, the following evaluation classes are used to characterize our results:
        \begin{description}
            \item[True Positive 'SURE':]\hfill \\This is the class of all correct matches, and in which the algorithm had a confident sense during the assignment. The goal is to have as many results as possible in this classification.
            \item[False Negative 'UNSURE':]\hfill \\In this class are all results where the algorithm led to the correct result, but the matching was classified with little confidence. Possibilities for this are either that the distances between users were so small that it is difficult to judge which one used the asset; or a distance estimate that deviated from the actual distance.
            \item[True Negative 'UNSURE':]\hfill \\This class represents the matches with a wrong result, but where the trust was also low. This class is the analog of the previous one, but where the matching was wrong.
            \item[False Positive 'SURE':]\hfill \\The last class is the one that can be described as definitely false. The algorithm has made a matching here, which is wrong - but was quite sure that the matching is correct. These are the results that are to be minimized.
        \end{description}
        
        \begin{figure}
          \centering
          \stackinset{l}{5pt}{b}{2pt}{(a)}{\confmat{{{347,143},{5,51}}}{\textbf{Matching}}{{"Correct","Wrong"}}{\textbf{Trustlevel}}{{"Sure", "Unsure"}}{2}{1.3}}
          \hspace{0.2cm}
          \stackinset{l}{5pt}{b}{2pt}{(b)}{\confmat{{{410,18},{0,6}}}{\textbf{Matching}}{{"Correct","Wrong"}}{\textbf{Trustlevel}}{{"Sure", "Unsure"}}{2}{1.3}}
          \caption{Confusion matrix of the (a) indoor and (b) outdoor experiments. Results reported in absolute numbers of measurements, dark blue color means many results, white means few.}
          \label{fig:confusion-matrix}
        \end{figure}
        The reported results can be seen in \cref{fig:confusion-matrix} and demonstrate the algorithm's performance in both indoor and outdoor scenarios.
        
        In the indoor environment, the algorithm achieved an overall accuracy of 89.7\%, with the accuracy defined in \cref{eq:accuracy}
        \begin{equation}\label{eq:accuracy}
            A = \frac{C}{T}
        \end{equation}
        where \(C\) is the number of correct matches and \(T\) is the total number of matches.

        Notably, the majority of correct classifications were made with high confidence, as evidenced by the "Guessed correct and 'SURE'" category, giving a recall as per \cref{eq:recall} of 70.8\%. 
        \begin{equation}\label{eq:recall}
            R = \frac{C_\text{sure}}{C}
        \end{equation}
        where \(C_\text{sure}\) is the number of correct matches with high trust level.

        The precision, indicating the probability that when the algorithm assigns a trust level of 'SURE', the matching was correct as defined in \cref{eq:precision}, in the indoor setting was exceptionally high (98.6\%). However, addressing the cases where the algorithm exhibited uncertainty is essential. 
        \begin{equation}\label{eq:precision}
            P = \frac{C_\text{sure}}{T_\text{sure}}
        \end{equation}
        where \(T_\text{sure}\) is the total number of matches with high trust level.

\begin{figure}[htpb!]
            \centering
            \includesvg[width=\columnwidth]{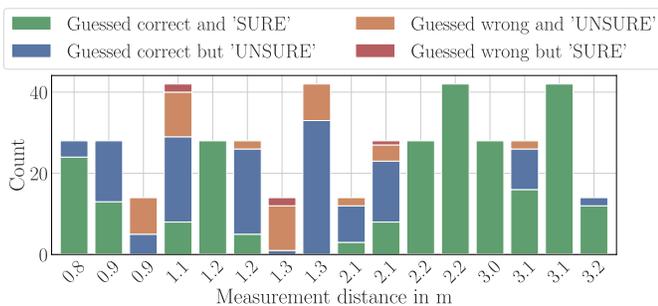}
            \vspace*{-5mm}
            \caption{Results in the indoor setup, divided up according the ground truth distance between asset and user. Does not include results of dynamic measurements.}
            \label{fig:results-confusion}
        \end{figure}

        In the "Guessed correct but trust level 'UNSURE'" category, there were 143 instances indoors, highlighting situations where the algorithm correctly identified the match but lacked confidence. \cref{fig:results-confusion} shows that a higher distance between the different workers generally resulted in a more reliable assignment between the worker and the tools. 

        In contrast, the outdoor scenario demonstrated an even higher overall accuracy of 98.6\%. The algorithm exhibited a remarkable recall of 95.8\% and perfect precision (100\%) in correctly classifying matches with high confidence ("Guessed correct and 'SURE'"). This suggests the algorithm's robustness in outdoor environments, where there are potentially fewer collisions and reflections, and therefore, a more deterministic path loss occurs.

        The absence of "Guessed wrong and 'SURE'" instances in the outdoor environment is promising, indicating that the algorithm, when wrong, tends to express uncertainty. This is a crucial characteristic, as it prevents the algorithm from confidently producing incorrect results.

        It should also be noted that a similarly high precision of 98.9\% was achieved in the dynamic scenarios, with the recall even increasing to 92.8\%. It can therefore be concluded that the detection of individual matchings works reliably even under more dynamic constraints. The reasons for the higher values could be the different dynamics between the correct matching and all other possible matchings: while there is relatively little movement between the user and his asset, the relative change in distance between a user and the other assets is greater - resulting in a more variable and bigger distance estimation.

        In conclusion, while the algorithm showcases strong performance in both indoor and outdoor settings, addressing instances of uncertainty, especially in indoor scenarios, can be a focus for improvement. These findings contribute valuable insights to the ongoing development and optimization of asset-matching algorithms in various environmental conditions.

    \subsection{Power measurements on the wearable device}
    \cref{tab:power_profile} shows a breakdown of the power consumption of the basic functionality and algorithms implemented on the wearable device. A working day of 8 hours was assumed, during which 20 different devices were active and recognized for 1.5 hours each. This corresponds to 15k advertisements that have to be processed by the wearable device per day.
    \begin{table}[htpb!]
            \centering
    	\caption{Power consumption breakdown of the wearable device (\textsc{EcoTrack}) during different activities. System voltage \qty{3.8}{\volt}}
    	\label{tab:power_profile}
    	\vspace{-0.3cm}
    	\renewcommand{\arraystretch}{1.3}
            \begin{center}
    	\begin{tabular}{@{}llrrr@{}}
    		\toprule
    		\multirow{2}{*}{\textbf{Part}}                              & \multirow{2}{*}{\textbf{Power state}}           & \textbf{Power}   & \textbf{Duration}& \textbf{Energy}  \\
    		\textbf{}                                        & \textbf{}           & \textbf{consumption}   & \textbf{per day}& \textbf{per day}  \\ \midrule
    		Baseline                                        & switched-off                   & \qty{296.00}{\nano\watt}        & \qty{16}{\hour}  & \qty{4.7}{\micro\watt{}\hour}\\ \midrule
    		\multirow{2}{*}{\acs{MCU}}                      & \acs{BLE} scan.                & \qty{38.38}{\milli\watt}     & \qty{8}{\hour}  & \qty{0.31}{\watt{}\hour}    \\
    		                                                & updating \acs{EKF}            & \qty{21.70}{\milli\watt}     & \qty{1.37}{\second}  & \qty{8.2}{\micro\watt{}\hour}    \\
    		\hline
    		\multirow{3}{*}{{\acs{LTE-M}}}                  & transmitting             & \qty{456.00}{\milli\watt}    & \(<\qty{10}{\second}\)  & \(<\qty{1.3}{\milli\watt{}\hour}\)  \\
    		                                                & idle connected                 & \qty{9.12}{\milli\watt}      & \qty{8}{\hour}  & \qty{73}{\milli\watt{}\hour} \\
    		                                                & deep sleep                     & \qty{266.00}{\micro\watt}    & \qty{16}{\hour} & \qty{4.3}{\milli\watt{}\hour}  \\
    		\midrule
    		\multicolumn{2}{@{}l}{\textbf{Average per day}}                          & \textbf{\qty{16.06}{\milli\watt}} & & \textbf{\qty{0.39}{\watt{}\hour}}    \\
    		        
    		\bottomrule
    	\end{tabular}
     \end{center}
    	\vspace{-0.3cm}
    \end{table}
    
    The largest energy consumer on the wearable device is the constant scanning for advertisements, with  \qty{0.31}{\watt{}\hour}. This is followed by the \ac{LTE-M} module, which is in cyclic idle/active mode most of the time (DRX = \qty{1.28}{\second}, no eDRX) and active for less than \qty{10}{\second} per day in order to transmit the collected data to the server. Compared to this power consumption, the energy consumption of the \ac{EKF} is small, with \qty{8.2}{\micro\watt{}\hour}. One iteration of the \ac{EKF} requires 5698 cycles on the \textsc{EcoTrack}, corresponding to an active time of \qty{1.37}{\second} for the 15k daily advertisements.   

    In total, the wearable device consumes an average power of \qty{16.06}{\milli\watt}. With the available \qty{10800}{\milli\watt{}\hour} battery, this results in a runtime of 28 days; enough to be charged only once every four weeks. This is particularly important, as it causes no substantial overhead to charge the devices.
\section{Conclusion}\label{sec:conclusion}
This study introduced a novel approach enabling asset-user matching in industrial environments using \acf{BLE}-enabled low-power \ac{IoT} devices. The system comprises \ac{BLE}-based tags equipped with accelerometers, wearable devices with internet access, and two algorithms: 
\begin{enumerate}[label=\roman*)]
\item A distance estimation algorithm based on \acf{EKF} to enhance the accuracy of distance estimations from \ac{RSSI} measurements. Despite challenges such as noise and measurement errors, the system achieved a median distance estimation error of \qty{0.49}{\meter} in the range of interest.
\item A cloud-based algorithm for asset-to-user matching that effectively matched assets with their users, achieving an accuracy of 87.7\% in indoor environments and 98.6\% in outdoor environments. 
\end{enumerate}
Furthermore, the system was validated through extensive indoor and outdoor experiments in real construction settings. This work proved that it is feasible to provide asset-user matches in real-world scenarios underscoring its utility for enhanced operational efficiency and safety. Additionally, the proposed approach optimizes power consumption across all components. The \ac{BLE} tags can operate for years on a single coin-cell battery. By reducing the data transmission load to the cloud server, the wearable devices are able to extend their operational lifetime to approximately 28 days on a \qty{10800}{\milli\watt{}\hour} battery. This aspect is crucial for widespread adoption in industrial environments where maintenance overhead must be minimized. Future works could focus on extending the testing to different environments with a sensitivity analysis on the algorithm's tuning parameters to validate its robustness and generalization capabilities.

\bibliographystyle{IEEEtran}
\bibliography{bib/IEEEabrv, bib/references}
\balance
\begin{IEEEbiography}[{\includegraphics[width=1in,height=1.25in,clip,keepaspectratio]{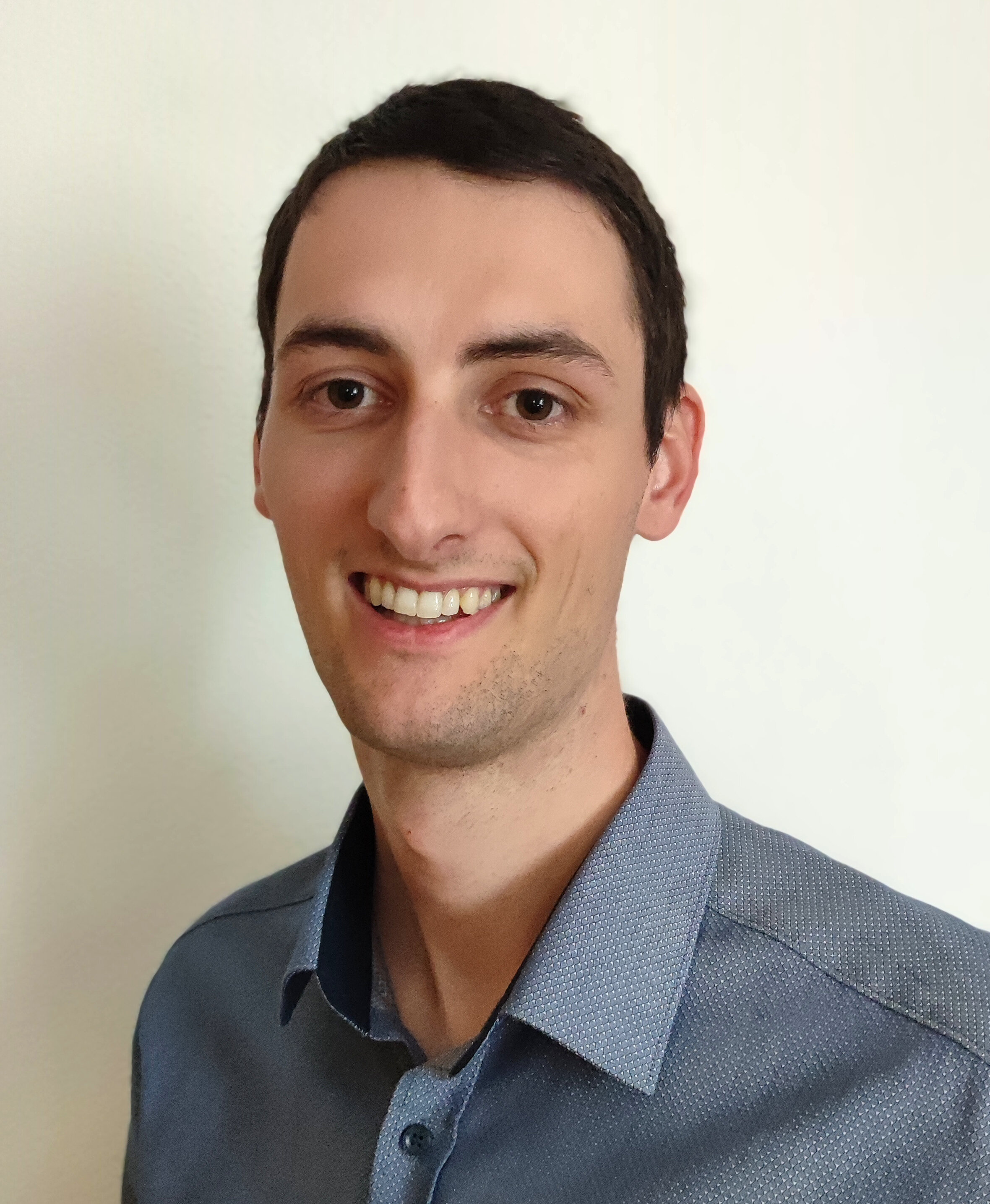}}]{Silvano Cortesi} (GS'22) received the B. Sc. and the M.Sc. degree in electronics engineering and information technology from ETH Zürich, Zürich, Switzerland in 2020 and 2021, respectively. He is currently pursuing the Ph.D. degree with the Center for Project-Based Learning at ETH Zürich, Zürich, Switzerland. His research work focuses on indoor localization, ultra-low power and self-sustainable IoT, wireless sensor networks and energy harvesting.
\end{IEEEbiography}

\begin{IEEEbiography}[{\includegraphics[width=1in,height=1.25in,clip,keepaspectratio]{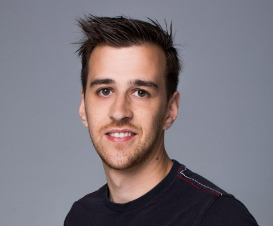}}]{Michele Crabolu} is an expert in biomedical embedded wearable technologies, sensor fusion and TinyML algorithms with a PhD in bioengineering and robotics from University of Genoa, Genoa, Italy, and more than ten clinical validation studies on wearable technologies. In 2021 he received a postgraduate diploma in Machine Learning and Artificial Intelligence from Columbia University. At the time of writing, he is technical program manager of the Corporate Research and Technology at Hilti Corporation. Before joining Hilti, he has worked for Xsens Technologies as Lead Engineer leading the research on sensor fusion and motion capture subjects.
\end{IEEEbiography}

\begin{IEEEbiography}[{\includegraphics[width=1in,height=1.25in,clip,keepaspectratio]{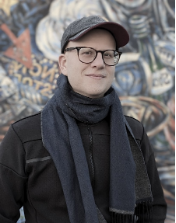}}]{Prodromos-Vasileios Mekikis} works as a Researcher at the Corporate Research and Technology unit of the Hilti Group. At Hilti, he is engaged in exploring innovative solutions for enhanced productivity, safety, and sustainability in the construction industry through the integration of cutting-edge IoT technologies and data-driven insights. In 2020, he was awarded a Marie Skłodowska-Curie Postdoctoral fellowship for his research on UAV networks. He got his PhD from the Technical University of Catalonia (UPC) in 2017. He also holds an Electrical and Computer Engineering degree (2010) and an M.Sc. in System-on-Chip design (2012) from Aristotle University of Thessaloniki (AUTH) and Royal Institute of Technology (KTH), respectively. His main research interests include connectivity in massive IoT networks, UAV-based networking, wireless power transfer, embedded systems design, and network function virtualization. 
\end{IEEEbiography}

\begin{IEEEbiography}[{\includegraphics[width=1in,height=1.25in,clip,keepaspectratio]{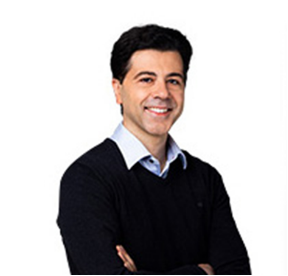}}]{Giovanni Bellusci} has broad R\&D industry experience in diverse technology domains including IoT, Machine Learning, Robotics, Sensing Technologies, and Sensor Fusion algorithms. He is currently Head of Robotics \& Mechatronics at Hilti Corporation, a global supplier of tools and services for the construction industry. Before joining Hilti, he was the Director of Technology \& CTO at Xsens Technologies, a leading innovator in motion capture and position tracking solutions for applications in Industrial, Robotics, Wearables, and Health. He holds a PhD Degree in Electrical Engineering from Delft University of Technology, the Netherlands, and an Executive MBA with Distinction from London Business School, UK.
\end{IEEEbiography}

\begin{IEEEbiography}[{\includegraphics[width=1in,height=1.25in,clip,keepaspectratio]{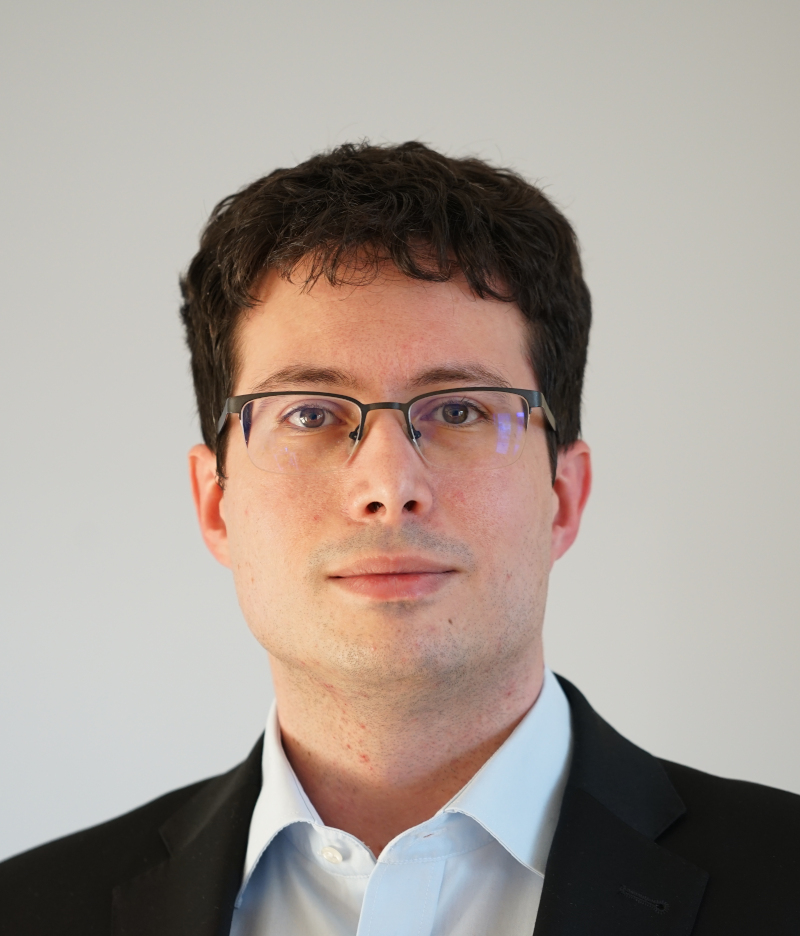}}]{Christian Vogt} (M'20) received the M.Sc. degree and the PhD in electrical engineering and information technology from ETH Zürich, Zürich, Switzerland, in 2013 and 2017, respectively. He is currently a post-doctoral researcher and lecturer at ETH Zürich, Zürich, Switzerland. His research work focuses on signal processing for low power applications, including field programmable gate arrays (FPGAs), IoT, wearables and autonomous unmanned vehicles.
\end{IEEEbiography}

\begin{IEEEbiography}[{\includegraphics[width=1in,height=1.25in,clip,keepaspectratio]{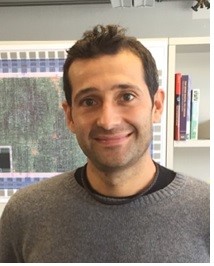}}]{Michele Magno} (SM'13) is currently a Senior Scientist at ETH Zürich, Switzerland, at the Department of Information Technology and Electrical Engineering (D-ITET). Since 2020, he is leading the D-ITET center for project-based learning at ETH. He received his master's and Ph.D. degrees in electronic engineering from the University of Bologna, Italy, in 2004 and 2010, respectively. He is working in ETH since 2013 and has become a visiting lecturer or professor at several universities, namely the University of Nice Sophia, France,  Enssat Lannion, France, Univerisity of Bologna and Mid University Sweden, where currently is a full visiting professor at the electrical engineering department. His current research interests include smart sensing, low-power machine learning, wireless sensor networks, wearable devices, energy harvesting, low-power management techniques, and extension of the lifetime of batteries-operating devices. He has authored more than 220 papers in international journals and conferences. He is a senior IEEE member and an ACM member. Some of his publications were awarded as best papers awards at IEEE conferences. He also received awards for industrial projects or patents.
\end{IEEEbiography}
\end{document}